\documentclass[12pt]{iopart}
\usepackage{graphicx}

\expandafter\let\csname equation*\endcsname\relax
\expandafter\let\csname endequation*\endcsname\relax
\usepackage{amsmath}
\usepackage{amssymb}
\usepackage{latexsym}
\usepackage{amsthm}
\usepackage{array}
\usepackage{amsfonts}
\usepackage{mathrsfs}
\usepackage{color}
\usepackage{verbatim}
\usepackage[caption=false]{subfig}
\usepackage{enumitem}
\usepackage{bbold}
\usepackage{ulem}
\usepackage[colorlinks=true,urlcolor=blue,citecolor=blue,linkcolor=blue]{hyperref}

\begin{document}
\title{Genuine monogamy relations in no-signaling theories --- a geometric approach}

\author{Junghee Ryu}
\address{National Institute of Supercomputing Division, Korea Institute of Science and Technology Information, Daejeon 34141, Republic of Korea}
\address{Centre for Quantum Technologies, National University of Singapore, 3 Science Drive 2, Singapore 117543}

\author{Daemin Lee}
\address{Department of Physics, Hanyang University, Seoul 04763, Republic of Korea}

\author{Jinhyoung Lee}
\address{Department of Physics, Hanyang University, Seoul 04763, Republic of Korea}

\author{Pawe\l{} Kurzy\'nski}
\address{Faculty of Physics, Adam Mickiewicz University, Umultowska 85, 61-614 Pozna\'n, Poland}
\address{Centre for Quantum Technologies, National University of Singapore, 3 Science Drive 2, Singapore 117543}

\author{Dagomir Kaszlikowski}
\address{Centre for Quantum Technologies, National University of Singapore, 3 Science Drive 2, Singapore 117543}
\address{Department of Physics, National University of Singapore, 2 Science Drive 3, Singapore 117551}

\date{\today}

\begin{abstract}	
Quantum correlations are subject to certain distribution rules represented by so-called monogamy relations. Derivation of monogamy relations for multipartite systems is a non-trivial problem, as the multipartite correlations reveal their behaviors in a way different from bipartite systems. We here show that simple geometric properties of probabilistic spaces, in conjunction with no-signaling principle, lead to {\it genuine} monogamy relations for a large class of Bell type inequalities for many qubits. The term of `genuine' implies that only one out of $N$ Bell inequalities exhibits a quantum violation. We also generalize our method to qudits. Using the similar geometric approach with a quasi-distance employed, we derive Svetlichny-Zohren-Gill type Bell inequalities for $d$-dimensional tripartite systems, and show their monogamous nature. 
\end{abstract}

\maketitle

\newcommand{\bra}[1]{\langle #1\vert} 
\newcommand{\ket}[1]{\vert {#1}\rangle} 
\newcommand{\abs}[1]{\vert#1\vert} 
\newcommand{\avg}[1]{\langle#1\rangle}
\newcommand{\braket}[2]{\langle{#1}|{#2}\rangle}
\newcommand{\commute}[2]{\left[{#1},{#2}\right]}
\newcommand{\mdo}[1]{\left[#1\right]}
\newcommand{\remark}[1]{\textcolor{blue}{#1}}
\newcommand{\del}[1]{\textcolor{red}{\sout{#1}}}
\newcommand{\add}[1]{\textcolor{blue}{\uwave{#1}}}
\newcommand{\sep}[1]{\left[ {#1}\right]}
\newcommand{\openone}{\mathbb {1}}

\newtheorem{theorem}{Theorem}

\section{Introduction}

Bell inequalities are a handy tool to check if spatially separated measurements have a local realistic description~\cite{Bell64,CHSH,ReviewBell}. As our universe is local, i.e., no perceived instantaneous action at a distance, any observable departure from local realism must be rather subtle---we do not directly observe the lack of local realism but only its consequences. One of them is the existence of strong quantum mechanical correlations (usually called non-local) and most of Bell inequalities relies on the differences between these correlations and the local realistic ones. The algorithm for Bell inequalities is deceptively simple: construct linear algebraic inequalities with correlation functions whose local realistic bounds are violated by the quantum correlations.

Initially, Bell inequalities were formulated for bipartite systems~\cite{Bell64,CHSH} and it was expected that for a large number of parties the system should loose its quantum character (non-locality) due to the correspondence principle~\cite{Bohr}. However, it was soon realized that multipartite systems exhibit an even more complex departure from the local realism~\cite{GHZ,MERMIN90}. Since then considerable efforts have been devoted to study multipartite systems in this context~\cite{ReviewBell,Pan2012}.

Multipartite systems exhibit another interesting property called correlation monogamy~\cite{Masanes06, Toner09, Ramanathan18, Tran18, Ramanathan14, Pawlowski09, Aolita12, Augusiak14, Augusiak17,Kurzynski11}. Monogamy imposes limits on the strength of distributed non-local correlations, i.e., the stronger the non-local correlations between two systems $A$ and $B$ are the weaker they are between the system $A$ and some other system $C$. It was shown that many known monogamies are a direct consequence of the no-signaling principle~\cite{Masanes06,Pawlowski09}---all involved parties $A, B$ and $C$ cannot exchange any information faster than the speed of light. This is, of course, strictly forbidden by the general relativity theory. Thus, the principle of no-signaling underpins monogamy of non-local correlations. Obviously, quantum mechanics is an example of a no-signaling theory and thus such monogamies are present in it.

The monogamy relations for the multipartite Bell inequalities were studied in~\cite{Ramanathan14, Pawlowski09, Aolita12, Augusiak14, Augusiak17}. They mainly focused on monogamies between bipartite divisions: a number of parties in some location $X$ is monogamous with the remaining parties at locations $Y$ and $Z$. In this limited scenario, the multipartite Bell inequalities are merely two-party Bell inequalities, each for two separated locations $X$ and $Y$ or $X$ and $Z$.

However, the quantum correlations in multipartite systems have specific traits that are not present in bipartite systems~\cite{HORODECKI}. Therefore, monogamies between more than two divisions, i.e., more than two multipartite Bell inequalities are interesting to study. This is a non-trivial problem. Indeed, it was reported that for four parties $A,B,C$ and $D$ one can find an entangled state such that three out of four possible tripartite Mermin type inequalities ($ABC$, $ABD$, $ACD$, and $BCD$) are violated. This is somewhat surprising as one would expect that only one inequality can be violated due to monogamy relations~\cite{Kurzynski11}.

In this paper we find {\it genuine monogamies} in the sense that only one Bell inequality can be violated regardless of the number of Bell inequalities involved. Because we only use the no-signaling principle, our results significantly limit the structure of quantum as well as any possible no-signaling correlations outside of the quantum theory (super-quantum correlations). Our method is based on simple geometric properties of probabilistic spaces and it leads to a series of new and genuine monogamy relations for any number of parties with dichotomic observables. As an illustration of the power of the method, we show that the `anomaly' reported in~\cite{Kurzynski11} vanishes, i.e., all possible tripartite Mermin type inequalities are genuinely monogamous. Our method also produces a genuinely tripartite Svetlichny-Zohren-Gill type Bell inequality for an arbitrary number of measurement outcomes and its monogamy. See Refs.~\cite{Aolita12, Augusiak17, Chen11, Bancal11} for $N$-partite Svetlichny inequalities with $d$ outcomes and the monogamy relations.

\section{Monogamy relations for many qubits}
One of the basic properties of any geometry is a distance $d(A,B)$ between two points $A$ and $B$. We introduce information-theoretic distances $d(A, B)$ defined on a space of probabilistic events such that $d(A, B)$ is a real function of the joint probabilities $p(A\cap B)$ of events $A$ and $B$. The distance $d(A,B)$ obeys the axioms of a metric: nonnegativity, symmetry, and most importantly for our applications, the triangle inequality. Note that the distance is valid for any sets of probabilistic events having a joint probability distribution. Therefore, if applied to some physical measurements $A$ and $B$, joint measurability of the corresponding physical properties (property $A$ and $B$) is implied.

It was shown that this geometric approach conveniently unifies different non-classical phenomena. It generates various kinds of bipartite Bell inequalities as well as some of the known tests of quantum contextuality~\cite{Schumacher91, Kurzynski14}. It also serves as a powerful tool to investigate correlation monogamies~\cite{Kurzynski14}. 

\subsection{Preliminaries}

The cornerstone of the method is a specific distance measure called statistical separation~\cite{Kolmogorov50,Renyi70,DUTTA18}.
Let us briefly discuss the statistical separation for two and three probabilistic events. We first define symmetric difference between two probabilistic events $A,B$ as $A \oplus B \equiv (A - B) \cup (B - A)$. A probability measure of the symmetric difference, $P(A \oplus B)$, is the statistical separation of the events $A$ and $B$. It can also be written as $P(A \oplus B)= P(A) + P(B) -2P(A \cap B)$. Note that the statistical separation conforms to all axioms of distance, including the triangle inequality: $P(A \oplus B)+P(B \oplus C) \geq P(A \oplus C)$. This can be derived using the following facts: (a) in the symmetric difference every event is its own inverse, i.e., $A \oplus A=\varnothing$, where $\varnothing$ is the empty set, (b) the operation $\oplus$ is commutative and associative, i.e.,  $A \oplus B = B \oplus A$, and $(A \oplus B) \oplus C =A \oplus (B \oplus C)$. The above properties together with the definition of the statistical separation give us: $(A \oplus B) \oplus (B \oplus C) =(A \oplus C)$ and $P(X) + P(Y) \geq P(X \oplus Y)$. The triangle inequality is obtained  by replacing $X$ with $A \oplus B$ and $Y$ with $B \oplus C$.

For the three event, one has $A \oplus B \oplus C \equiv (A \cap B \cap C) \cup (A \cap \overline{B} \cap \overline{C}) \cup (\overline{A} \cap B \cap \overline{C}) \cup (\overline{A} \cap \overline{B} \cap C)$, where $\overline{X}$ is the complement of $X$. All three events in the brackets are mutually exclusive so the statistical separation reads $P(A \oplus B \oplus C) = P(A \cap B \cap C) + P(A \cap \overline{B} \cap \overline{C}) + P(\overline{A} \cap B \cap \overline{C}) + P(\overline{A} \cap \overline{B} \cap C)$. 
We stress that the triangle inequality also holds for the statistical separation for three events, e.g., $P(A \oplus B \oplus C) + P(C \oplus D \oplus E) \geq P(A \oplus B \oplus D \oplus E)$. See \cite{DUTTA18} for the derivation of the triangle inequality and the generalization to $N$ events for the statistical separation. 

It is difficult to have an intuitive understanding of the statistical separation apart from thinking about it as a generalized ``geometric'' distance between three or more statistical events. This distance obeys the generalized triangle inequality that allows us to talk about ``geometric'' distance relations such as the ``length'' of the path between a certain chain of statistical events is longer or shorter than the other path. This intuition, as we will see in the rest of the paper, makes derivations of new multipartite Bell inequalities easier. To highlight this intuition we simplify the notation by putting, for example, $P(A \oplus B \oplus C) = \sep{A B C}$.


\subsection{Deriving Bell inequalities by using the triangle inequality}
We first consider a scenario when four parties $A, B, C$ and $D$ perform spatially separated (local) measurements of some physical property of their respective systems. If the observed property is found, the measurement yields $1$ as an outcome and $0$ otherwise. Each party randomly and independently selects one out of two measurement settings $i \in \{1,2\}$. The observer's $X$ detection event for the setting $i$ is denoted by $X_i$. This event happens with the probability $P(X_i)$. A joint probability for three events $A_i$, $B_j$, and $C_k$, which is the probabilistic measure of their intersection $A_i \cap B_j \cap C_k$, is denoted as $P(A_i, B_j, C_k) \equiv P(A_i \cap B_j \cap C_k)$.

Let us derive a Bell inequality from the statistical separation. Using the following triangle inequalities for the statistical separation 
\begin{eqnarray}
\sep{A_1B_2 C_2} +  \sep{A_2B_1 C_2} &\geq& \sep{A_1B_2A_2B_1}, \nonumber \\
\sep{A_1 B_2 A_2 B_1} + \sep{A_2B_2C_1} &\geq& \sep{A_1B_1C_1},
\label{eq:tri_local}
\end{eqnarray}
and adding them together, we arrive at the following Bell type inequality
\begin{eqnarray}
\mathcal{B}_{ABC} =\sep{A_1B_2C_2}+\sep{A_2B_1C_2} 
+\sep{ A_2B_2C_1} - \sep{A_1B_1C_1}\geq 0.
\label{eq:tri_bell_LR}
\end{eqnarray}
This separation Bell inequality is a generalization of the quadrangle inequality known in elementary geometry and first implemented to test local realism by Schumacher \cite{Schumacher91}.
The above tripartite Bell inequality is tight~\cite{DUTTA18}. This inequality is violated by quantum mechanics with a GHZ state (see Section \ref{APX:Quantum_violation}). The quantum violation is possible because in the derivation of~(\ref{eq:tri_bell_LR}) we assumed the existence of the statistical separation between the events corresponding to the measurements of non-commuting observables, $A_1, A_2$ and $B_1,B_2$ respectively. In any local realistic (LR) model such a statistical separation exists but the quantum violation of (\ref{eq:tri_bell_LR}) shows that it does not exist in quantum theory. This is in line with Fine's theorem~\cite{Fine} as the existence of the statistical distance between non-commuting observables (non-comeasurable events) would imply the existence of a joint probability distribution for non-commuting observables. For a more detailed explanation, see \ref{APX:Kolmogorov}.
Note that a different chaining of triangle inequalities lead to the Bell inequalities with a minus sign at a different separation. However, such all variant Bell inequalities are physically equivalent. In order to represent which separation is assigned minus, we denote the inequality in (\ref{eq:tri_bell_LR}) as $\mathcal{B}_{ABC}^{111}$. Additionally, we would like to mentioned that the lowerbound of the separation inequality for LR model is given zero as shown in (\ref{eq:tri_bell_LR}) and the one for the no-signaling principle is $-1$. The Bell inequalities with such boundaries are called normalized Bell inequalities, for example, see~\cite{Junge10, Rosset14} address normalized Bell functions that have the same LR and no-signaling models. However, our case is a natural result of deploying a chaining of triangle inequalities. We have no intention of deriving the normalized Bell inequalities.        
\begin{figure}[t]
	\centering
	\includegraphics[width=3cm]{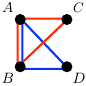}
	\caption{(color online). No-signaling monogamy diagram for two tripartite Bell inequalities $\mathcal{B}_{ABC}^{111}$ (red) and $\mathcal{B}_{ABD}^{122}$ (blue). Only one of the inequalities can be violated by quantum theory.}
	\label{fig:tripartite}
\end{figure}

\subsection{Primary monogamy relation}
Consider another Bell inequality of $\mathcal{B}_{ABD}^{111}$ which can be simply obtained by changing $C \rightarrow D$ in (\ref{eq:tri_bell_LR}). Simple adding two Bell inequalities does not result in a monogamy relation. Note that every LR model trivially satisfies the monogamy inequality $\mathcal{B}_{ABC}^{111} + \mathcal{B}_{ABD}^{111} \geq 0$.
This is becasue each term $\mathcal{B}_{ABC}^{111}$ and $\mathcal{B}_{ABD}^{111}$ is always non-negative for any LR model. 
 
We now derive a monogamy relation for the violations of Bell inequalities in any no-signaling theory. It is a non-trivial observation because there are no-signaling theories that can violate Bell inequalities. Quantum mechanics is an example of such a theory for which we can have $\mathcal{B}_{ABC}\leq 0$ and $\mathcal{B}_{ABD}\geq 0$ or vice versa. Our claim is that for all no-signaling theories we have
\begin{eqnarray}
\mathcal{M}_{AB;CD}\equiv \mathcal{B}_{ABC}^{111} + \mathcal{B}_{ABD}^{122} \underset{\mathcal{NS}}{\geq} 0.
\label{EQ:MERMIN_SEP}
\end{eqnarray}
$\mathcal{M}_{AB;CD}$ is called the {\it primary} monogamy (see figure~\ref{fig:tripartite}).

The gist of our reasoning is to show that a suitable chaining of the triangle inequalities satisfied by any no-signaling theory leads to $\mathcal{B}_{ABC}$ and $\mathcal{B}_{ABD}$. The following set of triangle inequalities hold,
\begin{eqnarray}
\sep{A_2B_1C_2} + \sep{A_2B_1D_2} &\geq& \sep{C_2D_2}, \nonumber \\
\sep{C_2D_2} + \sep{A_1B_2C_2} &\geq& \sep{A_1B_2D_2}, \nonumber \\
\sep{A_2B_2C_1} + \sep{A_2B_2D_1} &\geq& \sep{C_1D_1}, \nonumber \\
\sep{C_1D_1} + \sep{A_1B_1D_1} &\geq& \sep{A_1B_1C_1}.
\label{eq:tri_mono}
\end{eqnarray}
These triangle inequalities are also derived from the mathematical properties of the statistical separation and the symmetric difference: (a) $P(X) + P(Y) \geq P(X \oplus Y)$ and (b) $(A \oplus B) \oplus (B \oplus C) =(A \oplus C)$. For instance, the first inequality is derived by replacing the event $X$ by the event $A_1B_2C_2$, the event $Y$ by the event $A_1B_2D_2$. Then the event $X\oplus Y$, because of the property (b), is equivalent to the event $(A_2B_1C_2)\oplus (A_2B_1D_2)=C_2D_2$. Combining this with the property (a) we recover the first triangle inequality. All the other inequalities in the above inequalities are derived in a similar way.  We would like to remark that these inequalities hold in any no-signalling model not only in a LR model. We also show how to derive these inequalities in a non-local model supplemented with the no-signalling condition. See \ref{APX:Kolmogorov} for more details.

Each triangle inequality has the common statistical separations $C_{i} D_{i}$ with $i \in \{1,2\}$. The no-signaling principle guarantees that any given separation is independent on the context it was measured with; the separation $C_{2} D_{2}$ in the first triangle inequality is the same as the separation in the second one. Without the no-signaling principle these two separations could be context dependent: $C_2D_2$ in the first inequality dependent on the context $A_2B_1$ and $C_2D_2$ in the second inequality dependent on the context $A_1B_2$. This common statistical separation $C_{i} D_{i}$ in~(\ref{eq:tri_mono}) cancels out when all triangle inequalities are added, resulting in the monogamy relation~(\ref{EQ:MERMIN_SEP}): $\sep{A_1B_2C_2} + \sep{A_2B_1C_2} + \sep{A_2B_2C_1} - \sep{A_1B_1C_1}-\sep{A_1B_2D_2} + \sep{A_2B_1D_2}  +\sep{A_2B_2D_1}  + \sep{A_1B_1D_1} \geq 0$.

Note that the minus signs appear in two Bell functions $\mathcal{B}_{ABC}^{111}$ and $\mathcal{B}_{ABD}^{122}$ at certain positions that are direct consequences of the separations' geometric properties encoded in~(\ref{eq:tri_mono}). As we will see later, this simple observation has strong consequences that make our monogamies different and stronger from those reported in~\cite{Masanes06, Toner09, Ramanathan18, Tran18, Ramanathan14, Pawlowski09, Aolita12, Augusiak14, Augusiak17, Kurzynski11}. To be more precise, we could put the minus sign in the $\mathcal{B}_{ABD}$ in front of any other separation without changing the monogamy and the physics of the problem. However, this innocent change leads to non-genuine no-signaling monogamies for more than two Bell functions as we will show below. The primary monogamy in (\ref{EQ:MERMIN_SEP}) will be used as a base method to derive genuine monogamy relations.

\begin{figure}[t]
	\centering
	\subfloat[]{
	\includegraphics[width=3cm]{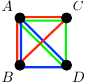}
	\label{FIG:3bellmono}
	}\quad \quad
	\subfloat[]{
	\includegraphics[width=3cm]{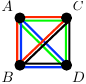}
	\label{FIG:fullmono}
	}
	\caption{(color online). Genuine monogamies for four tripartite Bell inequalities of $\mathcal{B}_{ABC}^{111}$ (red), $\mathcal{B}_{ABD}^{122}$ (blue), $\mathcal{B}_{ACD}^{212}$ (green), and $\mathcal{B}_{BCD}^{122}$ (black). (a) Genuine monogamy relation states that only one of the inequalities $\mathcal{B}_{ABC}^{111}, \mathcal{B}_{ABD}^{122}$, and $\mathcal{B}_{ACD}^{212}$ can be violated. (b) Full symmetric monogamy relation holds if we add one more Bell inequality $\mathcal{B}_{BCD}^{122}$.}
	\label{fig:4party}
\end{figure}

\subsection{Genuine monogamies of four tripartite Bell inequalities}
Let us now derive genuine monogamy relations involving many Bell inequalities such that only one of the inequalities can show the quantum violations. Figure~\ref{FIG:3bellmono} depicts the geometry of three tripartite Bell inequalities: $\mathcal{B}_{ABC}^{111}$ (red), $\mathcal{B}_{ABD}^{122}$ (blue), and $\mathcal{B}_{ACD}^{212}$ (green). This geometry implies the following monogamy:
\begin{eqnarray}
\mathcal{B}_{ABC}^{111}  + \mathcal{B}_{ABD}^{122} + \mathcal{B}_{ACD}^{212} \underset{\mathcal{NS}}{\geq} 0,
\label{EQ:MONO_4}
\end{eqnarray}
where each inequality reads
\begin{eqnarray}
\mathcal{B}_{ABC}^{111} &=&  \sep{A_1B_2C_2} + \sep{A_2B_1C_2} + \sep{A_2B_2C_1} - \sep{A_1B_1C_1}, \nonumber \\
\mathcal{B}_{ABD}^{122} &=& - \sep{A_1B_2D_2} + \sep{A_2B_1D_2} + \sep{A_2B_2D_1} + \sep{A_1B_1D_1}, \nonumber \\
\mathcal{B}_{ACD}^{212} &=&  \sep{A_1C_2D_2} - \sep{A_2C_1D_2} + \sep{A_2C_2D_1} + \sep{A_1C_1D_1}.
\label{EQ:three_Bell}
\end{eqnarray}
The proof follows an observation that any pair of the Bell inequalities in~(\ref{EQ:MONO_4}) is the primary monogamy. More precisely, for the Bell functions $\mathcal{B}_{ABC}^{111}$ and $\mathcal{B}_{ACD}^{212}$, the following triangle inequalities lead the primary monogamy as $ \mathcal{M}_{AC;BD} \equiv \mathcal{B}_{ABC}^{111} + \mathcal{B}_{ACD}^{212} \geq 0$:
\begin{eqnarray}
\sep{A_1B_2C_2} + \sep{A_1C_2D_2} &\geq& \sep{B_2D_2}, \nonumber \\
\sep{B_2D_2} + \sep{A_2B_2C_1} &\geq& \sep{A_2C_1D_2}, \nonumber \\
\sep{A_2B_1C_2} + \sep{A_2C_2D_1} &\geq& \sep{B_1D_1}, \nonumber \\
\sep{B_1D_1} + \sep{A_1C_1D_1} &\geq& \sep{A_1B_1C_1}.
\label{eq:ABC_ACD_tri}
\end{eqnarray}
For the primary monogamy $\mathcal{M}_{AD;BC} \equiv \mathcal{B}_{ABD}^{122} + \mathcal{B}_{ACD}^{212} \geq 0$, we have
\begin{eqnarray}
\sep{A_1 B_1 D_1} + \sep{A_1 C_1 D_1} &\geq& \sep{B_1 C_1}, \nonumber \\
\sep{B_1 C_1} + \sep{A_2 B_1 D_2} &\geq& \sep{A_2 C_1 D_2}, \nonumber \\
\sep{A_2 B_2 D_1} + \sep{A_2 C_2 D_1} &\geq& \sep{B_2 C_2}, \nonumber \\
\sep{B_2 C_2} + \sep{A_1 C_2 D_2} &\geq& \sep{A_1 B_2 D_2}.
\label{eq:ABC_BCD_tri}
\end{eqnarray}

The primary monogamies $\mathcal{M}_{AB;CD}, \mathcal{M}_{AC;BD}, \mathcal{M}_{AD;BC}$ are greater or equal to zero and thus $\mathcal{M}_{AB;CD}+\mathcal{M}_{AC;BD}+\mathcal{M}_{AD;BC}\geq 0$, which is exactly the formula~(\ref{EQ:MONO_4}). This new monogamy is genuine in the sense that only one of the three Bell functions can be negative, leaving the other two compatible with LR.
More precisely, if one Bell inequality in (\ref{EQ:MONO_4}), e.g., $\mathcal{B}_{ABC}^{111}$ shows a violation (it is negative), then the other inequalities, $\mathcal{B}_{ABD}^{122}$ and $\mathcal{B}_{ACD}^{212}$, should be positive (no violation) because the $\mathcal{B}_{ABC}^{111}$ must be monogamous with $\mathcal{B}_{ABD}^{122}$ and $\mathcal{B}_{ACD}^{212}$. This works for any other Bell inequalities in (\ref{EQ:MONO_4}).

Now we take a step further, throw in one more Bell inequality $\mathcal{B}_{BCD}^{122}$ (see figure \ref{FIG:fullmono}) and use the similar reasoning to prove that
\begin{eqnarray}
\mathcal{B}_{ABC}^{111} + \mathcal{B}_{ABD}^{122} + \mathcal{B}_{ACD}^{212} + \mathcal{B}_{BCD}^{122} \underset{\mathcal{NS}}{\geq} 0,
\label{EQ:full_mono}
\end{eqnarray}
where the $\mathcal{B}_{BCD}^{122}$ reads
\begin{eqnarray*}
\mathcal{B}_{BCD}^{122} &=&  -\sep{B_1C_2D_2} +\sep{B_2C_1D_2} + \sep{B_2C_2D_1} + \sep{B_1C_1D_1}.
\end{eqnarray*}
Similar to the previous case, one can construct the primary monogamies for $\mathcal{B}_{BCD}^{122}$ with the rest of Bell inequalities $\mathcal{B}_{ABC}^{111}, \mathcal{B}_{ABD}^{122}, \mathcal{B}_{ACD}^{212}$, respectively.

This is also the genuine monogamy in the same way as before, i.e., only one Bell inequality can be violated by no-signaling correlations. In contrast, the \cite{Kurzynski11} derives a ``Mermin monogamy'' consisting of four Mermin inequalities such that three of them can be simultaneously violated by a four-qubit partially entangled state. Each inequality reads $E_{122}+E_{212}+E_{221}-E_{111}\leq 2$. Here, $E_{ijk}$ stands for the usual correlation function of the measurement results corresponding to the settings $ijk$. These correlation functions can be cast in the form of separation used in this paper, see for instance \cite{DUTTA18}. Note that all Mermin inequalities in their monogamy assign the minus sign to the correlation functions $E_{111}$. This leads to the simultaneous violations of up to three inequalities and thus to a non-genuine monogamy. This can be easily `fixed' by changing the position of the minus sign in the inequalities---a fix that is not necessary in our method. Genuine monogamies are definitely more desirable in quantum communication protocols such as cryptography, secret key sharing etc.~\cite{Gisin02, Prabhu12, Giorgi11, Kumar16}. They also can be used in characterization of quantum many body systems as well as in quantum biology~\cite{Chandran07, Song13, Qin16, Zhu12, Chanda14}.

\subsection{Generalization to multipartite case}
We extend our monogamy relations for a general case of $N$-partite Bell inequalities. For the binary outcomes, we introduce two sets of symmetric differences for $N$-party measurement events: $\mathcal{X}=\{A_1 \oplus B_2 \oplus \dots \oplus N_2$, and all cyclic permutations$\}$ and $\mathcal{Y}=\{A_1 \oplus B_1 \oplus \dots \oplus N_1\}$ for odd $N$, and for even $N$ one more term $A_2 \oplus B_2 \oplus \dots \oplus N_2$ is added to the set $\mathcal{X}$. In the set $\mathcal{X}$, each local measurement event with the setting $2$ appears an even number of times because of the cyclic permutations, so that these terms can be dropped out in deriving the $N$-partite separation Bell inequalities. This is because every event in the symmetric difference is its own inverse, i.e., $X \oplus X=\varnothing$. Thus, the additional term of symmetric difference $A_2 \oplus B_2\oplus  \dots \oplus N_2$ is needed for even number of parties. Thus, the $N$-partite inequality reads
\begin{eqnarray}
\mathcal{B}_{AB\dots N}=\sum_{i} P(\mathcal{X}_i) - P(\mathcal{Y}_1) \geq 0,
\label{eq:Nbellineq}
\end{eqnarray}
where the $\mathcal{X}_i$ ($\mathcal{Y}_1$) implies the $i$th ($1$st) element of the set $\mathcal{X}$ ($\mathcal{Y}$).
For example of $N=4$, the Bell function reads
\begin{eqnarray}
\mathcal{B}_{ABCD}^{1111} &=& [A_{1} B_{2} C_{2} D_{2}] + [A_{2} B_{1} C_{2} D_{2}] + [A_{2} B_{2} C_{1} D_{2}] \nonumber \\
&+& [A_{2} B_{2} C_{2} D_{1}] + [A_{2} B_{2} C_{2} D_{2}] - [A_{1} B_{1} C_{1} D_{1}],
\label{eq:exampleN4}
\end{eqnarray}
and for $N=5$, 
\begin{eqnarray}
\mathcal{B}_{ABCDE}^{11111} &=& [A_{1} B_{2} C_{2} D_{2} E_{2}] + [A_{2} B_{1} C_{2} D_{2} E_{2}] + [A_{2} B_{2} C_{1} D_{2} E_{2}] \nonumber \\
&+& [A_{2} B_{2} C_{2} D_{1} E_{2}] + [A_{2} B_{2} C_{2} D_{2} E_{1}] - [A_{1} B_{1} C_{1} D_{1} E_{1}],
\label{eq:exampleN5}
\end{eqnarray}
where we use the notation as $P(A \oplus B \oplus C) = \sep{A B C}$. We can see that the Bell function for $N=4$ (even $N$) in (\ref{eq:exampleN4}) has an additional separation $[A_{2} B_{2} C_{2} D_{2}]$ as we explain above.
Note that such a geometric inequality is invariant with respect to swapping of any one separation from the set $\mathcal{X}$ and the other from $\mathcal{Y}$. That is, all variants of Bell inequalities are equivalent. 

The above $N$-partite separation Bell inequality is different from those of Mermin-Ardehali-Belinski-Klyshko, known as MABK inequality~\cite{MERMIN90, MABK}. It is a kind of a chained Bell inequality which has been studied in~\cite{Santos86, Pykacz91, Zukowski14}.
We show the quantum violations in Section \ref{APX:Quantum_violation}.

With another party denoted by $N'$, we have the following no-signaling primary monogamy: $\mathcal{B}_{A\dots N-1 N} + \mathcal{B}_{A\dots N-1N'} \geq 0$. Similar to the tripartite case, one can group the separations in the monogamy into two sets such that each of which has one separation with the minus sign. Then, as before, under the assumption of no-signaling, each set can be rewritten into two triangle inequalities linked via the context-independent separations [see the argument below (\ref{eq:tri_mono})]. Moreover, by using the primary monogamy we can derive the genuine monogamies for $N$-partite system: Any four $N$-partite Bell inequalities out of $(N+1)$ inequalities must hold the no-signaling monogamy relations. For $N=5$, one example of the genuine monogamy is $\mathcal{B}_{ABCDE}^{11111} + \mathcal{B}_{ABCDF}^{22122} + \mathcal{B}_{ABCEF}^{22212} + \mathcal{B}_{ABDEF}^{22122} \geq 0$. Each Bell inequality is given in a form of (\ref{eq:Nbellineq}). The separations with minus sign for each inequality are $-\sep{A_1B_1C_1D_1E_1}, -\sep{A_2B_2C_1D_2F_2}, -\sep{A_2B_2C_2E_1F_2}$, and $-\sep{A_2B_2D_1E_2F_2}$. This is an example of an $AB$ division monogamy in the sense that the parties $A$ and $B$ are in the monogamous relations with the remaining parties as $C, D, E, F$, i.e., $AB$ is monogamous with $CDE$, $CDF$, $CEF$, and $DEF$. Note that because of the symmetries presented in our approach this division holds for any two parties. For any $N$, our results can be extended to a $(N-3)$ division monogamy relations. A rule of assigning the minus sign to separations is as follows: for a given division, the minus is given to the separations in which the measurement settings for the remaining parties are $\{111,122,212,122\}$ (refer to the measurement settings for $C, D, E, F$ of the $N=5$ example). Unlike the tripartite scenario, we cannot derive a fully symmetric monogamy relations for arbitrary $N$ where all Bell inequalities are involved. This is in contrast to the monogamy~(\ref{EQ:full_mono}) and  we conjecture that this can be improved if we consider the Bell inequalities with more separations or more than two measurement settings.

\section{$d$-outcome scenario}
An extension for an arbitrary number of measurement outcomes $d$ requires a use of a quasi-distance---a metric that is not symmetric. Interestingly, our method still stands with a few simple modifications to account for the lack of symmetry.

\subsection{Quasi-distance}
Let us first show the triangle inequality for the quais-distance. For two events, it reads
\begin{eqnarray}
P(A>B) + P(B>C) \geq P(A>C),
\label{apx:eq_quasi_tri}
\end{eqnarray}
where $P(X>Y) \equiv \sum_{x} \sum_{y<x} P(X^x, Y^y)$. A joint probability $P(X^x, Y^y)$ describes that two parties obtain the outcomes $x$ and $y$, respectively. To prove this, define the event $A \leq B \equiv \bigcup_{a \leq b} A^{a} \cap B^{b}$. Note that $(A \leq B) \cap (B \leq C) \subseteq (A \leq C)$ and $(A > B) \cup (B > C) \supseteq (A > C)$. These result in the triangle inequality in~(\ref{apx:eq_quasi_tri}). The event $X>Y$ is not symmetric so the order of events in the triangle inequality is crucial.

We define a quasi-distance between events $A,B,C$ as
\begin{eqnarray}
P([A+B]<C) \equiv \sum_{a,b} \sum_{c>[a+b]} P(A^a, B^b, C^c).
\label{eq:quasi_prob}
\end{eqnarray}
Here $P(A^a,B^b,C^c)$ is the probability of a joint event where three parties detect the outcomes $a, b$, and $c$, respectively, and $[x]$ reads `$x$ modulo $d$' \cite{ACIN05} (note that the notation $[\cdot]$  is different to the one previously used to present the statistical separation). The quasi-distance satisfies all axioms of a regular distance {\it sans} symmetry, importantly the triangle inequality: $P([A+B]<C) + P(C<[D+E]) \geq P([A+B]<[D+E])$. 

\subsection{Bell inequalities by using the quasi-distance}
Consider a tripartite scenario involving two choices of measurement settings $X_{i}$ $(i=1, 2)$ each of which has $d$ possible outcomes: $X_{i}=0, 1, \dots, d-1$. Using the properties of the quasi-distance we have the following triangle inequalities:
\begin{eqnarray}
\fl P([A_1 + B_1] < C_2) + P(C_2 < [A_2 + B_1]) \, &\geq& P([A_1 + B_1] < [A_2 + B_1]), \nonumber \\
\fl P([A_1 + B_1] < [A_2 + B_1]) + P([A_2 + B_1] < C_1) \,&\geq& P([A_1 + B_1] < C_1), \nonumber \\
\fl P(C_2 < [A_1+B_2])+ P([A_1+B_2] < C_1) \,&\geq& P( C_2 < C_1), \nonumber \\
\fl P( C_2 < C_1) + P( C_1 < [A_2+B_2]) \,&\geq& P( C_2 < [A_2+B_2]).
\end{eqnarray}
Adding these triangle inequalities lead to the Bell inequality as
\begin{eqnarray}
\fl &&\mathcal{B}_{ABC}=P([A_1+B_1] < C_2) + P(C_2 < [A_2+B_1]) + P([A_2+B_1] < C_1) -P([A_1+B_1] < C_1) \nonumber \\
\fl &+& P(C_2 < [A_1+B_2])+ P([A_1+B_2] < C_1) + P( C_1 < [A_2+B_2]) - P( C_2 < [A_2+B_2]) \geq 0.
\label{EQ:3_ZG}
\end{eqnarray}
To this end, we followed exactly the steps outlined in the derivation of (\ref{eq:tri_local}). Nothing has changed because those steps do not rely on the symmetry of the common statistical separation.


Another Bell inequality $\mathcal{B}_{ABD}$ can be derived by the swaps of the type $C_i \rightarrow D_i$ and $``<" \rightarrow ``>"$. These swaps depend on the sign and their logic is easier to understand from the inequalities shown below rather than to describe in words. The triangle inequalities read
\begin{eqnarray}
\fl P(D_2 < [A_1 + B_1]) + P([A_1 + B_1] < D_1) \, &\geq& P(D_2 < D_1), \nonumber \\
\fl P(D_2 < D_1) + P(D_1 < [A_2 + B_1]) \,&\geq& P(D_2 < [A_2 + B_1]), \nonumber \\
\fl P([A_1+B_2] < D_2 )+ P(D_2 < [A_2+B_2] ) \,&\geq& P( [A_1+B_2]  < [A_2+B_2] ), \nonumber \\
\fl P( [A_1+B_2]  < [A_2+B_2] ) + P( [A_2+B_2]  < D_1) \,&\geq& P( [A_1+B_2]  < D_1).
\end{eqnarray}

We highlight that the Bell inequality~(\ref{EQ:3_ZG}) is the Svetlichny-type extension of the Zohren-Gill inequality~\cite{Zohren08} (see e.g., Refs.~\cite{Aolita12, Augusiak17, Chen11, Bancal11} for the multipartite and $d$ outcomes extensions of the Svetlichny inequalities). It is a genuinely tripartite Bell inequality for $d$-outcomes to detect the genuinely tripartite nonlocality (see \ref{APX:proof_genuine} for the proof). The quantum violations of the inequality imply that the given quantum state is the genuinely tripartite entanglement. It is violated by the generalized GHZ state (see Section \ref{APX:Quantum_violation}). By a simple swap $``[A+B] <C" \rightarrow ``[A+B]>C"$ we get yet another Bell inequality. 

\subsection{Monogamy relations}
By following exactly the same route as in the tripartite scenario discussed before we arrive at the primary monogamy
\begin{eqnarray}
\mathcal{B}_{ABC} + \mathcal{B}_{ABD} \underset{\mathcal{NS}}{\geq} 0.
\label{EQ:3_ZG_mono}
\end{eqnarray}
The following chain of triangle inequalities gives the monogamy relation:
\begin{eqnarray}
P(D_2 < [A_1 + B_1]) + P([A_1 + B_1] < C_2) &\geq& P([D_2 < C_2]), \nonumber \\
P([D_2 < C_2]) + P(C_2 < [A_2 + B_1]) &\geq& P(D_2 < [A_2 + B_1] ), \nonumber \\
P(D_1 < [A_2 + B_1]) + P([A_2 + B_1] < C_1) &\geq& P([D_1 < C_1]), \nonumber \\
P([D_1 < C_1]) + P([A_1 + B_1] < D_1) &\geq& P([A_1 + B_1] < C_1), \nonumber \\
P(C_2 < [A_1 + B_2]) + P([A_1 + B_2] < D_2) &\geq& P([C_2 < D_2]), \nonumber \\
P([C_2 < D_2]) + P(D_2 < [A_2 + B_2]) &\geq& P(C_2 < [A_2 + B_2] ), \nonumber \\
P(C_1 < [A_2 + B_2]) + P([A_2 + B_2] < D_1) &\geq& P([C_1 < D_1]), \nonumber \\
P([C_1 < D_1]) + P([A_1 + B_2] < C_1) &\geq& P([A_1 + B_2] < D_1).
\label{apx:quasi_triangle}
\end{eqnarray}
For an arbitrary $d$, the genuine monogamy conditions for tripartite systems and their extension to multipartite cases are still unknown, and they will be discussed elsewhere. It is worth noticing that Ref.~\cite{Augusiak17} also presents the monogamy relations for $N$-partite Svetlichny inequalities for $M$ measurement settings and $d$ outcomes. In particular, for $N=3$ and $M=2$ the monogamy relation is similar in form to (\ref{EQ:3_ZG_mono}).

\section{Quantum violations}
\label{APX:Quantum_violation}
We show the quantum violations of the Bell inequalities presented in the paper. The violations can be observed for the $N$-qudit GHZ state
\begin{eqnarray}
\ket{\Psi}=\frac{1}{\sqrt{d}} \sum_{n=0}^{d-1} \ket{n}^{\otimes N}.
\label{EQ:NDGHZ}
\end{eqnarray}
As stated previously, spatially separated parties measure two projectors on the GHZ state. 

First consider the separation Bell inequality for the binary outcomes. For odd $N$, the Bell inequality in (\ref{eq:Nbellineq}) reads
\begin{eqnarray}
&&P(A_1\oplus B_2\oplus \cdots \oplus N_2) + \mbox {(its cyclic permutations)} \nonumber \\
&-& P(A_1 \oplus B_1 \oplus \cdots \oplus N_1) \geq 0.
\label{APX_eq:sep_bell}
\end{eqnarray}
For even $N$, one more separation $P(A_2\oplus B_2\oplus \cdots \oplus N_2)$ is added to the inequality. For the $N$-qubit GHZ state, i.e., $d=2$ in~(\ref{EQ:NDGHZ}), the separation is given by
\begin{eqnarray}
P(A \oplus B \oplus \cdots \oplus N) = \frac{1}{2} \left( 1+ k \left<\bigotimes_{s=1}^{N} \hat{b}_s \cdot \vec{\sigma}\right> \right),
\label{APX_eq:prob}
\end{eqnarray}
where $\hat{b}_s$ are Bloch vectors of the local projectors for the observer $s$. The variable $k$ has two possible values depending on $N$; $k=-1$ for even $N$, and $k=+1$ for odd $N$. The symbol $\left< \cdot \right>$ denotes the expectation value for the GHZ state.

We show the proof of (\ref{APX_eq:prob}) for $N=2$ before generalizing to arbitrary $N$. The projector for the measurement setting $i$ of the party $A$ is given by
\begin{eqnarray}
\hat{\Pi}_{A_i} = \frac{1}{2} \left( 1 + \hat{b}_{A_i} \cdot \vec{\sigma} \right),
\label{APX_eq:proj}
\end{eqnarray}
where $\hat{b}_{A_i}$ is a unit Bloch vector for the $i$th setting of $A$. By definition of the statistical separation we have
\begin{eqnarray}
P(A_i \oplus B_j) &=& P(A_i) + P(B_j) - 2 P(A_i, B_j) \nonumber \\
&=& \text{Tr} \varrho (\hat{\Pi}_{A_i} \otimes \openone + \openone \otimes \hat{\Pi}_{B_j} - 2 \hat{\Pi}_{A_i} \otimes \hat{\Pi}_{B_j}),
\label{APX_eq:prob-op}
\end{eqnarray}
where $\varrho$ is a density operator for the given state. Using (\ref{APX_eq:proj}), the separation (\ref{APX_eq:prob-op}) leads to
\begin{eqnarray}
P(A_i \oplus B_j)= \text{Tr} \left[ \frac{1}{2} \left(\openone \otimes \openone - \hat{b}_{A_i} \cdot \vec{\sigma} \otimes \hat{b}_{B_j} \cdot \vec{\sigma}  \right) \varrho \right].
\end{eqnarray}
It can be extended to an arbitrary $N$ recursively (see Section $3$ of \cite{DUTTA18}). Note that the sign of the correlation measurement depends on the number of $N$ as $-1$ for even $N$ and $+1$ for odd $N$. $\qedhere$

To calculate (\ref{APX_eq:prob}), let us confine $\hat{b}_s$ to the $x$-$y$ plane with $b_{sz}=0$. We introduce complex variables $b_s = b_{sx} + i b_{sy}$, where $b_{sx}$ and $b_{sy}$ are $x$ and $y$ components of the local Bloch vectors $\hat{b}_s$, respectively. Then, for arbitrary $N$ we have
\begin{eqnarray}
\left<\bigotimes_{s=1}^{N} \hat{b}_s \cdot \vec{\sigma}\right> =\mathfrak{R}\left(\prod_{s} b_s\right),
\end{eqnarray}
where $\mathfrak{R}(b)$ is the real value of $b$. It is clear that the right-hand side removes imaginary terms, containing the even number of $b_y$'s. Letting $b_{sx} = \cos \theta_s$ and $b_{sy} = \sin \theta_s$, the complex variable $b_s = \exp (i\theta_s)$ so that
\begin{eqnarray}
\mathfrak{R}\left(\prod_{s} b_s\right)=\cos \theta,
\end{eqnarray}
where $\theta=\sum_{s} \theta_s$. Then, the separation reads
\begin{eqnarray}
P(A \oplus B \oplus \cdots \oplus N) &=& \frac{1}{2} (1+ k \cos \theta) \nonumber \\
&=& \begin{cases} 
(1+k)/2, & \mbox{if } \theta=2m\pi \\ 
(1-k)/2, & \mbox{if } \theta=(2m+1)\pi
\end{cases}
\label{EQ:case}
\end{eqnarray}
for an integer $m$. This is the perfect correlation for the GHZ state. Therefore, when we set the measurement setting $1$ as $x$ and the setting $2$ as $x$ rotated around $z$ by $\pi/(N-1)$ for each respective party, we will have that $P(A_1\oplus B_2\oplus \cdots \oplus N_2)$ and its cyclic permutations are zero. But, as $P(A_1 \oplus B_1 \oplus \cdots \oplus N_1)=1$, finally the inequality (\ref{APX_eq:sep_bell}) leads to $-1 \ge 0$ for odd $N$. For an even $N$, we set the measurement setting $1$ as $x$ rotated around $z$ by $\pi/N$ and the setting $2$ as $x$ rotated around $z$ by $-\pi/N(N-1)$ for each respective party. Such settings show that the $P(A_1\oplus B_2\oplus \cdots \oplus N_2)$ and its cyclic permutations are zero, and $P(A_1 \oplus B_1 \oplus \cdots \oplus N_1)=1$. Additional term $P(A_2\oplus B_2 \oplus \cdots \oplus N_2) = [1- \cos ( -\pi/(N-1) ) ]/2$ is not perfectly correlated, but it is always less than one. Thus, we still observe the violation (a weaker one).

\begin{figure}[t]
	\centering
	\includegraphics[width=8cm]{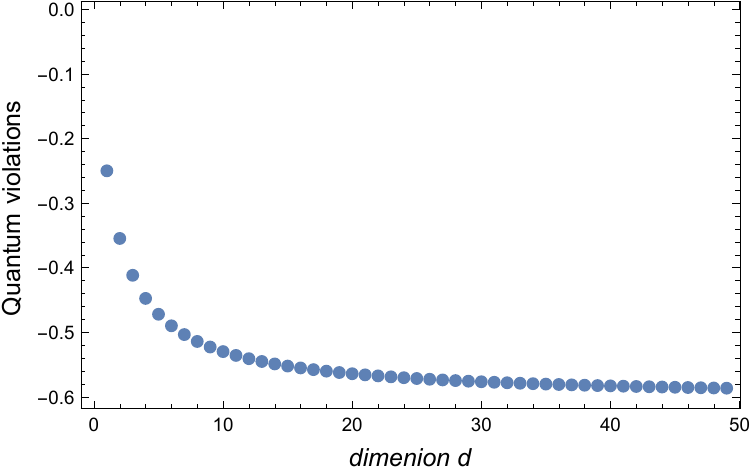}
	\caption{Value of the left-hand side of the Bell inequality in (\ref{EQ:3_ZG}) as a function of the dimension $d$ up to $50$. As the lowerbound of the inequality by the local realistic description is zero, thus the negative values of the results imply the quantum violations.}
	\label{FIG:violation}
\end{figure}

For higher dimensional cases, we used the quasi-distance to derive Bell inequalities with two local measurements. To see the quantum violations, we use a Fourier basis
\begin{eqnarray}
\ket{x}_{X_i} &=& \frac{1}{\sqrt{d}} \sum_{n=0}^{d-1} \omega^{n(x+\varphi_i)} \ket{n},
\end{eqnarray}
where $\omega=\exp(2 \pi i/d)$ and $\varphi_i$ is a local phase of the $X_i$th measurement.
Then, the probability $P(A^{a}_{i}, B^{b}_{j}, C^{c}_{k})$ to get the outcomes $a,b,c$ given the settings $i,j,k$ for a GHZ state~(\ref{EQ:NDGHZ}) reads 
\begin{eqnarray}
P(A^{a}_{i}, B^{b}_{j}, C^{c}_{k}) = \frac{1}{d^4} \left| \sum_{n=0}^{d-1} \omega^{n(a_i+b_j - c_k  + \varphi_{ijk})} \right|^2,
\end{eqnarray}
where $\varphi_{ijk} = \varphi_{i} +\varphi_{j}-\varphi_{k}$. To violate the Bell inequality, we set $\varphi_{111}=1$, $\varphi_{222}=-1$, $\varphi_{112}=\varphi_{211}=\varphi_{121}=1/3$, and $\varphi_{212}=\varphi_{122}=\varphi_{221}=-1/3$. Figure~\ref{FIG:violation} shows the quantum predictions of the left-hand side in (\ref{EQ:3_ZG}) up to $d=50$. Negative values imply the quantum violations.

\section{Conclusions}
We presented a new way of deriving the monogamies of the Bell inequalities violations.
Our method is based on geometric properties of probability spaces called statistical separation, first introduced by Kolmogorov and developed further in the context of non-classical correlations in \cite{Kolmogorov50,Renyi70,DUTTA18}. The cornerstone of results obtained is the triangle inequality.

We derived new monogamies for $N$ parties, each performing measurement of two dichotomic observables. These monogamies are stronger than any other monogamies of this kind reported in the literature \footnote{The term `strong' is used to highlight that our monogamy relations involving many Bell inequalities are genuine such that only one Bell inequality can be violated by the quantum theory and by any no-signaling theory.}. They are stronger because of their strictly exclusivity, i.e., only one out of $N$ Bell inequalities can be violated. This can have potential applications in various quantum information protocols. It also improves our understanding of the relation between multipartite quantum, super-quantum correlations and no-signaling principle.    

Using the similar geometric approach we also derived a genuinely tripartite Svetlichny-Zohren-Gill type Bell inequality for two $d$-outcome measurements and showed its quantum violation. We also showed a no-signaling monogamy for this inequality. Interestingly, we had to use the quasi-distance in these derivations. As far as we know, this is the first usage of this particular quasi-distance in the context of non-classical correlations. It would be interesting to see how to extend this to multipartite correlations and how to derive genuine multipartite monogamies. This will be presented in a forthcoming work.

\ack
We thank the anonymous referee for bring the Bell inequality in \cite{Aolita12, Augusiak17} to our attention and for valuable comments. This research was supported by the National Research Foundation, Prime Minister’s Office, Singapore and the Ministry of Education, Singapore under the Research Centres of Excellence programme and Singapore Ministry of Education Academic Research Fund Tier 3 (Grant No. MOE2012-T3-1-009). This research was also supported through the National Research Foundation of Korea (NRF) grant (No. 2019R1A2C2005504, No. 2019M3E4A1079666, and NRF-2020M3E4A1079792), funded by the Korea government(MSIT).

\appendix
\section{Kolmogorov probabilistic description}
\label{APX:Kolmogorov}
Here we describe the event and its probability measure in terms of Kolmogorov probability space denoted by $(\Omega, \mathcal{F}, P)$, where $\Omega$ is a sample space, $\mathcal{F}$ an event space, $P$ a probability measure. Consider the simplest scenario in which two observers, Alice and Bob perform two binary local measurements each.

We first show the local realistic description for probabilistic events and then extend it to the nonlocal model. In general, the nonlocal model includes not only no-signaling theories but also signaling models. Then, we show how the no-signaling condition applies to the nonlocal events and how it yields the desired triangle inequalities.

\subsection{Local model}
\label{app:localmodel}
In a local realistic model, all possible outcomes form a sample space which can be written as
\begin{equation}
\Omega_{\mathcal{L}} = \{ (a_1, a_2, b_1, b_2) \: | \: a_i, b_j = 0, 1\},
\end{equation}
for all $i, j$. The total number of elements in the sample space $\Omega_{\mathcal{L}}$ is $2^4 = 16$ (in a general case of $n$ observers, $m$ local settings, and $d$ outcomes, it is $d^{n m}$). Event space $\mathcal{F}$ is a set of subsets of the sample space $\Omega_{\mathcal{L}}$. An event $X$ is an element of ${\cal F}$. For instance, the event with outcome $\alpha$ for Alice's choice of the measurement setting $i=1$ is given by
\begin{eqnarray}
A_{i=1}^{\alpha} = \{ (a_1 = \alpha, a_2, b_1, b_2) \: | \: a_2, b_1, b_2 = 0, 1 \}.
\label{EQ:event_A}
\end{eqnarray}
Alice's detection event is encoded by $\alpha=1$ and no detection event by $\alpha=0$. The events $A_{i}^{0}$ and $A_{i}^{1}$ are subsets of the sample space $\Omega_{\mathcal{L}}$. Similarly, we can define the events $A_2^\alpha, B_1^\beta$, and $B_2^\beta$. For a given probability measure $P$, event $X$ has probability 
\begin{equation}
P(X) = \sum_{\omega \in \Omega} \xi_{X} (\omega) P(\omega),
\label{EQ:prob_event}
\end{equation}
where $\omega$ is a sample point, that is, an element of the sample space and $\xi_X(\omega)$ is an indicator function with values $1$ if $\omega \in X$ and $0$ otherwise. Obviously, it holds $P(A_{i}^{0})+P(A_{i}^{1})=1$ for all $i$. In the main text, as we deal with the `detection' events only, $\alpha$s are omitted unless there is a possibility of confusion.

\subsection{Nonlocal model}
\label{app:tri_mono}

Now let us discuss a nonlocal model. While measurement outcomes in a local model only depend on the local settings, measurement outcomes in a nonlocal model can be affected by the measurement settings of the spatially separated parties. For example, Alice's measurement outcome $a$ depends on both her and Bob's settings, $i$ and $j$ respectively, and thus is denoted by $a_{ij}$. 

The sample space for a nonlocal model is given by
\begin{equation}
\Omega_{\mathcal{NL}} = \{ (a_{11}, a_{12}, a_{21}, a_{22}, b_{11}, b_{12}, b_{21}, b_{22}) \: | \: a_{ij}, b_{ij} = 0,1\},
\label{eq:NLmodel}
\end{equation}
for all $i,j$.
Each element of the event space can be denoted by
\begin{equation}
A^{\alpha}_{ij} = \{ (\cdots, a_{ij} = \alpha, \cdots) \: | \: a_{i' j'}, b_{ij} = 0,1 \},
\end{equation}
for $i \neq i'$ and $j \neq j'$. The probability of the event has the same formula as in (\ref{EQ:prob_event}) except that in the nonlocal model the sample is given by (\ref{eq:NLmodel}).

We are now interested in no-signaling theories. It means that the marginals of the joint probability distribution are independent from the other party's measurement setting. For a joint probability distribution $P(a,b | x,y)$ this means
\begin{eqnarray}
\sum_{b} P(a,b | x,y) = \sum_{b} P(a,b | x,y')  \quad \forall a, x, y, y',
\label{EQ:nosingal_cond}
\end{eqnarray}
where $a$ and $b$ are Alice's and Bob's measurement outcomes for the settings $x$ and $y$, respectively. 
In our case, this condition can be rewritten as
\begin{eqnarray*}
P(a,0|1,1) + P(a,1|1,1) = P(a,0|1,2) + P(a,1|1,2).
\end{eqnarray*}

The no-signaling condition in terms of the probability measure can be written as
\begin{equation}
P(A_{ij}^{a}) = P(A_{ij'}^{a}).
\label{EQ:prob_nosignal}
\end{equation}
One can see immediately that this probability doesn't depend on the Bob's measurement settings. The no-signaling condition for Bob is defined in a similar way.
The formulas of (\ref{EQ:nosingal_cond}) and (\ref{EQ:prob_nosignal}) are thus related
\begin{eqnarray}
P(a,0|1,1) = \sum_{\vec{a}, \vec{b}} P(a_{11}=a, \vec{a}, b_{11}=0, \vec{b}).
\end{eqnarray}
Here $\vec{a} = (a_{12}, a_{21}, a_{22})$, $\vec{b}=(b_{12}, b_{21}, b_{22})$ and we have  $P(a,0|1,1) + P(a,1|1,1) = P(A_{11}^{a})$.

\subsection{Proof of the triangle inequalities in (\ref{eq:tri_mono})}
\label{app:proof_Kol}
Here we prove the triangle inequalities~(\ref{eq:tri_mono}) presented in the main text using the Kolmogorov probability description explained in the previous sections. As four parties are involved, the event for each party has four subscripts. For example, Alice's detection event is denoted by $A_{ijkl}$, where $i$ is Alice's measurement setting and $j, k, l$ are the measurement settings chosen by others. Then, the triangle inequalities in~(\ref{eq:tri_mono}) can be rewritten as 
\begin{eqnarray}
\sep{A_{2122} B_{2122} C_{2122}} + \sep{A_{2122} B_{2122} D_{2122}} &\geq& \sep{C_{2122} D_{2122}} = \sep{C_{1222} D_{1222}}, \nonumber \\ 
\sep{C_{1222} D_{1222}} + \sep{A_{1222} B_{1222} C_{1222}} &\geq& \sep{A_{2122} B_{2122} D_{2122}}, \nonumber \\
\sep{A_{2211} B_{2211} C_{2211}} + \sep{A_{2211} B_{2211} D_{2211}} &\geq& \sep{C_{2211} D_{2211}} =  \sep{C_{1111} D_{1111}}, \nonumber \\ 
\sep{C_{1111} D_{1111}} + \sep{A_{1111} B_{1111} D_{1111}} &\geq& \sep{A_{1111} B_{1111} C_{1111}},
\end{eqnarray}
where we used the no-signaling condition and the following properties of the common separation: $\sep{C_{1222} D_{1222}}=\sep{C_{2122} D_{2122}}$ and $\sep{C_{2211} D_{2211}} = \sep{C_{1111} D_{1111}}$. More precisely, the first two subscripts for $C$ and $D$ imply the choices of the measurement settings by the parties $A$ and $B$, respectively. Because the parties are spatially separated, the choices of the measurement settings for the parties $A$ and $B$ do not affect the choices made by  $C$ and $D$ (no-signaling condition). Similarly, this condition also applies to $C$ and $D$. All these observations yield the triangle inequalities (\ref{eq:tri_mono}) in the main text.

\section{Proof of genuine tripartite Bell inequality}
\label{APX:proof_genuine}

Here we show that the Bell inequality (\ref{EQ:3_ZG}) in the main text is the genuine tripartite Bell inequality. A violation of such inequality implies that the systems present the genuine tripartite entanglement. It was first discovered by Svetlichny for tripartite case~\cite{Svetlichny87} and later generalized to multipartite one~\cite{Aolita12, Chen11, Bancal11}. 

To this end, we convert the probability functions in (\ref{EQ:3_ZG}) to averages of floor functions. This was used in \cite{Tavakoli16} to prove that Zohren-Gill inequality is equivalent to CGLMP inequality for two qudits. For $a, b \in \{0,1,...,d-1\}$, the relation reads $P(B<A)=-\langle\lfloor\frac{B-A}{d}\rfloor\rangle$, where $\lfloor \cdot \rfloor$ is the floor function and $\langle f(A,B) \rangle = \sum_{a,b} f(a,b) P(A^a,B^b)$. As 
\begin{eqnarray}
-\frac{d-1}{d}\leq\frac{b-a}{d}\leq\frac{d-1}{d},
\nonumber
\end{eqnarray}
we have
\begin{eqnarray}
\lfloor\frac{b-a}{d}\rfloor=
\begin{cases} 0 & \mbox{for } b \geq a \\
-1 & \mbox{for } b<a.
\nonumber
\end{cases}
\end{eqnarray}
We generalize the method to three qudits and we get the relation $P([A+B]<C)=-\langle\lfloor\frac{[A+B]-C}{d}\rfloor\rangle$. Note that by definition of $[x]$ the range reads $[a+b] \in \{0,1,...,d-1\}$, thus we can apply the above relation to our tripartite Bell inequality [see (\ref{eq:quasi_prob}) in the main text for the definition of the probability function for tripartite case].

The method enables to rewrite the Bell inequality (\ref{EQ:3_ZG}) in a form of
\begin{eqnarray}
\fl&-&\langle\lfloor\frac{A_1+B_1-C_2}{d}\rfloor\rangle
-\langle\lfloor\frac{-A_2-B_1+C_2}{d}\rfloor\rangle
-\langle\lfloor\frac{A_2+B_1-C_1}{d}\rfloor\rangle
+\langle\lfloor\frac{A_1+B_1-C_1}{d}\rfloor\rangle
\nonumber \\
\fl&-&\langle\lfloor\frac{-A_1-B_2+C_2}{d}\rfloor\rangle
-\langle\lfloor\frac{A_1+B_2-C_1}{d}\rfloor\rangle
-\langle\lfloor\frac{-A_2-B_2+C_1}{d}\rfloor\rangle
+\langle\lfloor\frac{-A_2-B_2+C_2}{d}\rfloor\rangle
\geq0, \nonumber \\
\fl
\label{EQ:3_ZG_appendix_floor}
\end{eqnarray}
where we used the following properties: $[x]=x-d\lfloor\frac{x}{d}\rfloor$, $\lfloor x+n \rfloor = \lfloor x \rfloor +n$ for integer $n$, and $\lfloor \lfloor x \rfloor\rfloor = \lfloor x \rfloor$. We show that our inequality (\ref{EQ:3_ZG_appendix_floor}) is equivalent to the genuine tripartitie Bell inequalities in \cite{Aolita12, Bancal11}. Note that $\langle\lfloor\frac{[A+B]-C}{d}\rfloor\rangle=\langle\lfloor\frac{A+B-C}{d}\rfloor\rangle-\langle\lfloor\frac{A+B}{d}\rfloor\rangle$ and the terms of pair variables in a form of $\langle\lfloor\frac{A+B}{d}\rfloor\rangle$ are cancelled out. Multiplying (\ref{EQ:3_ZG_appendix_floor}) by $d$, using the relation $[x]=x-d\lfloor\frac{x}{d}\rfloor$ and applying the relation $[-x]=(d-1)-[x-1]$ we rewrite the inequality (\ref{EQ:3_ZG_appendix_floor}) in another form of
\begin{eqnarray}
\fl &&\langle[A_1+B_1-C_2]\rangle
+\langle[-A_2-B_1+C_2]\rangle
+\langle[A_2+B_1-C_1]\rangle
+\langle[-A_1-B_1+C_1-1]\rangle
\nonumber \\
\fl &+&\langle[-A_1-B_2+C_2]\rangle
+\langle[A_1+B_2-C_1]\rangle
+\langle[-A_2-B_2+C_1]\rangle
+\langle[A_2+B_2-C_2-1]\rangle
\geq2(d-1). \nonumber \\
\fl
\label{EQ:3_ZG_appendix_absolute}
\end{eqnarray}
This form of our inequality coincides with the genuine tripartite Bell inequality $S_{3,d}$ in \cite{Bancal11} when the sign of $C$ is flipped to be plus.
Moreover, applying $B_{1} \rightarrow -B_{2}$, $B_{2} \rightarrow -B_{1}$ and $C_i \rightarrow -C_i$ in (\ref{EQ:3_ZG_appendix_absolute}) leads to another genuine tripartite Bell inequality $I_{M}^{3}$ for two measurement settings ($M=2$) in \cite{Aolita12}.

It is worth noting that the inequality (\ref{EQ:3_ZG}) remains genuinely tripartite under party permutations. When the parties $A$ and $C$ are exchanged, we have
\begin{eqnarray}
\fl&&P([B_1+C_1]<A_2)+P(A_2<[B_1+C_2])+P([B_1+C_2]<A_1)-P([B_1+C_1]<A_1) + \nonumber \\
\fl&&P(A_2<[B_2+C_1])+P([B_2+C_1]<A_1)+P(A_1<[B_2+C_2])-P(A_2<[B_2+C_2]) \geq 0. \nonumber \\
\fl
\label{ZG_appendix_BC}
\end{eqnarray}
By applying the similar procedure from (\ref{EQ:3_ZG_appendix_floor})-(\ref{EQ:3_ZG_appendix_absolute}), the inequality (\ref{ZG_appendix_BC}) is expressed in a form of (\ref{EQ:3_ZG_appendix_absolute}) but flipping signs of $A$ and $C$. Note that such a sign flip for the local party leaves unchanged physical properties of the Bell inequality. We have the similar equivalence also for exchanging $B$ and $C$.


\end{document}